\documentclass[aps,twocolumn,showpacs,preprintnumbers,amsmath,amssymb]{revtex4}

\usepackage{epsfig}
\usepackage{graphicx}
\usepackage{amssymb} 
\usepackage{here}
\usepackage{eucal}
\usepackage{dcolumn}
\usepackage{bm}

\def\cut{\Lambda}

\def\T{{\mathcal T}}

\newcommand{\beq}{\begin{equation}}
\newcommand{\eeq}{\end{equation}}
\newcommand{\be}{\begin{equation}}
\newcommand{\ee}{\end{equation}}
\newcommand{\beqa}{\begin{eqnarray}}
\newcommand{\eeqa}{\end{eqnarray}}
\newcommand{\ba}{\begin{array}}
\newcommand{\ea}{\end{array}}

\def\R{{\cal R}}
\def\T{{\cal T}}

\def\End{\end{document}}

\def\to{\rightarrow}

\def\dis{\displaystyle}
\def\f{\frac}
\def\ov{\overline}

\def\[{\left[}
\def\]{\right]}
\def\({\left(}
\def\){\right)}

\def\under{\underline}

\def\U1EM{U(1)_{\rm em}}

\def\R{\mathcal R}

\def\leqq{\leqslant}
\def\geqq{\geqslant}

\def\d{\delta}

\def\[{\left[}
\def\]{\right]}
\def\dis{\displaystyle}

\def\cut{\Lambda}

\def\Eu{E^{\star}}

\def\nL{\nu_L^{~}}

\def\fbar{\bar{f}}
\def\leqq{\leqslant}
\def\geqq{\geqslant}

\def\under{\underline}

\begin{document}

\preprint{\large hep-ph/0502178}


\title{Scales of Mass Generation for Quarks, Leptons
       and Majorana Neutrinos
      }


%
\author{{\sc Duane A. Dicus} ~and~ {\sc Hong-Jian He}\,\,}
\affiliation{Center for Particle Physics, 
             University of Texas at Austin, TX 78712, USA
}




\begin{abstract}
\noindent
We study  \,$2\to n$\, inelastic fermion-(anti)fermion scattering
into multiple longitudinal weak gauge bosons and derive
universal upper bounds on the scales of
fermion mass generation by imposing  unitarity of the $S$-matrix.
We place new upper limits on the scales
of fermion mass generation, independent of the electroweak
symmetry breaking scale.
We find that the strongest $2\to n$ limits fall in a narrow range, 
$3-170$\,TeV (with $n=2-24$),   
depending on the observed fermion masses. 
\end{abstract}

\pacs{
      12.15.Ff, 14.60.Pq, 11.80.-m, 12.60.-i \hfill 
      Phys. Rev. Lett. (2005), in Press.}

%


\maketitle


\noindent  
\under{\it New Puzzle for Scales of Fermion Mass Generation~}
\vspace*{3mm}

Understanding scales of mass generation for 
quarks, leptons and Majorana neutrinos poses a 
great challenge in elementary particle physics.
Mass generation in the standard model (SM)
relies on a single hypothetical Higgs boson.
Although the masses of weak bosons $(W^\pm,Z^0)$ only involve
electroweak gauge couplings $(g,\,g')$ and the vacuum expectation
value $v=(\sqrt{2}G_F)^{-\f{1}{2}}$,  the quark/lepton masses
arise from products of Yukawa couplings $y_f^{~}$ and $v$\,.\, 
Contrary to the gauge interactions, the Yukawa couplings $y_f^{~}$ are  
flavor-dependent and completely arbitrary,
exhibiting a huge empirical hierarchy 
\,$y_e^{~}/y_t^{~}=m_e^{~}/m_t^{~} \simeq 3\times10^{-6}$\,
between the electron and the top quark.
No compelling principle requires the fermion mass generation to
share the same mechanism as the $W/Z$ gauge bosons.
Furthermore, the tiny neutrino masses 
$\,0.05\,{\rm eV}\lesssim m_\nu^{~}\lesssim 1\,$eV\,\cite{superK}
cannot be generated by such a Higgs boson without
losing renormalizability\,\cite{wein5} or extending the SM
particle spectrum\,\cite{nu-seesaw,nu-rad}.
So far, no Higgs boson has been found, nor is any
Yukawa coupling experimentally measured ---  
the origin and scale of
fermion mass generation remain completely unknown.

What is wrong with just {\it putting all the bare masses into 
the SM Lagrangian by hand?}  
Such bare mass-terms can be made 
gauge-invariant in the nonlinear realization\,\cite{CCWZ},
but are manifestly nonrenormalizable. 
This causes unitarity violation in
high energy scattering at a scale $\Eu$. 
Generically, we define the scale 
$\cut_x$ for generating a mass $m_x^{~}$ 
to be the {\it minimal energy
above which the bare mass term for $m_x^{~}$ has to be replaced 
by a renormalizable interaction} (adding at least one new physical
state to the particle spectrum). Hence, the unitarity violation
scale $\Eu$ puts a {\it universal upper limit on $\cut_x$,}
\,$\cut_x \leqq \Eu$.

A bare mass-term for $(W^\pm ,Z^0)$ will cause 
the high energy \,$2\to 2$\, scattering of 
longitudinal weak bosons
to violate unitarity at a scale\,\cite{DM},
%
$\,E^\star_W ~\simeq~ \sqrt{8\pi}v ~\simeq~ 1.2\,{\rm TeV}\,.$\,
%
This puts an upper limit on the scale of 
electroweak symmetry breaking (EWSB) and 
justifies the TeV energy scale for the 
Large Hadron Collider (LHC) at CERN. 
Similarly, by adding bare mass terms for Dirac fermions 
an upper bound on the scale of fermion mass generation 
can be derived from the \,$2\to 2$\, inelastic scattering  
$f\bar{f} \to V_L^{a_1}V_L^{a_2}$ 
($V^a=W^\pm,Z^0$) \cite{AC,MVW},
\beq
\label{eq:UBf22}
E_f^\star ~\simeq~ \dis  \f{8\pi v^2}{~\sqrt{N_c}\,m_f^{~}~} \,,
\eeq
where $N_c = 3\,(1)$ for quarks (leptons).
Since this limit is proportional to 
$1/m_f^{~}$, it is independent of the  
bound $E^\star_W$ on the EWSB scale.
For all the SM fermions (except the top quark), 
the limit (\ref{eq:UBf22})
is substantially higher than $E^\star_W$.
For the scale of mass generation for Majorana neutrinos, 
Refs.\,\cite{scott-PRL,scott} derived
an analogous $2\to 2$ bound from the scattering
\,$\nL\nL\to V_L^{a_1}V_L^{a_2}$,  
\beq
\label{eq:UBnu22}
E_\nu^\star ~\simeq~ \dis  
\,4\pi v^2/m_{\nu}\,,
\eeq
which is around \,$10^{16}$\,GeV\, (the GUT scale) for a
typical neutrino mass \,$m_\nu^{~}\sim 0.05$\,eV \cite{superK}.

However, for $2\to n$ inelastic scattering,
$f\bar{f},\,\nL\nL$ $\to nV_L^a$ ($n\geqq 3$), 
Ref.\,\cite{scott} found that the $n$-body phase space
integration contributes a large energy factor, $s^{n-2}$,
to further enhance the cross section 
(in addition to the energy dependence $s^1$ 
from the squared amplitude), where $\sqrt{s}$ is the center of 
mass energy. From the unitarity condition on the total
cross section\,\cite{scott},
\beq
\label{eq:UC2n}
\sigma_{\rm inel}[2\to n] ~\leqq~  4\pi/s \,,
\eeq
the unitarity limit for 
$f~(\nL)$ would behave as\,\cite{scott}
\beq
\label{eq:UBfscot}
E_{f,\nu}^\star \sim \dis 
v\({v}/{\,m_{f,\nu}^{~}}\)^{\f{1}{n-1}}
\longrightarrow v\,,~~~({\rm as}\,~n\to{\rm large}),
\eeq
which could be pushed arbitrarily close to the weak scale $v$ and
thus become independent of the fermion mass 
$m_{f,\nu}^{~}$ for large enough \,$n$\,.\,
This raises a striking question: 
{\it is there an independent new scale for 
fermion mass generation revealed from the fermion-(anti)fermion
scattering into weak bosons?}  
We find the behavior (\ref{eq:UBfscot}) very counter-intuitive
since the kinematic condition forces
any \,$2\to n$\, unitarity limit $\Eu$
to grow at least linearly with \,$n$\,,
\beq
\label{eq:KC}
\dis
\sqrt{s} ~\,>~\, n M_{W(Z)} \,\simeq\, \dis\f{n}{3}v ~,
~~~\Longrightarrow~~~
\f{\,E^\star}{v} ~\,>~\, \f{n}{3} ~,
\eeq
which contradicts (\ref{eq:UBfscot}), 
posing a {\it new puzzle.}

\begin{figure*}
\label{fig:ff-npi}
\begin{center}
\vspace*{-5mm}
\hspace*{-3mm}
\includegraphics[width=18cm,height=6cm]{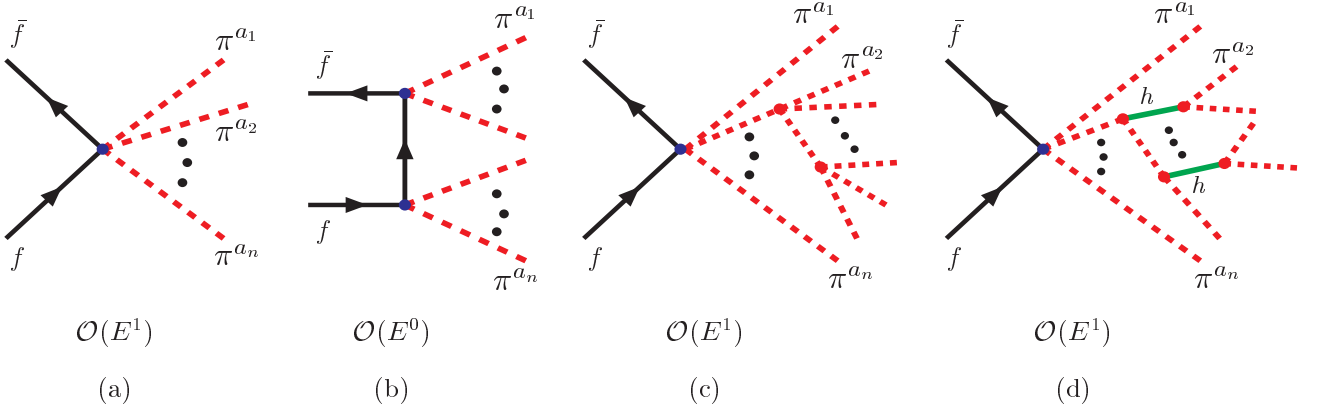}
\vspace*{-6mm}
\caption{Typical contributions to
$f\fbar (\nL\nL) \to n\pi^a$
scattering: 
(a) leading ``contact'' diagrams of ${\mathcal O}(E^1)$;
(b) sub-leading ``$t(u)$-channel'' diagrams of
${\mathcal O}(E^0)$  which can be ignored for the unitarity analysis;
(c) ``hybrid''  contact diagrams ($n\geqq 3$) including Goldstone
self-interactions from the EWSB sector;
(d) ``hybrid''  contact diagrams ($n\geqq 3$) including Goldstone
interactions with the EWSB quanta (illustrated here with the SM Higgs
boson $h^0$  as the simplest example of EWSB).
}
\vspace*{-0.5cm}
\end{center}
\end{figure*}

\vspace*{3mm}
\noindent  
\under{\it Resolution of the New Puzzle~}
\vspace*{2mm}

Consider high energy $2\to n$ inelastic scattering 
$\,f\fbar,$ $\nL\nL\to nV_L^a\,$. 
Its amplitude can be derived from the corresponding scattering 
$\,f\ov{f},\nL\nL\to n\pi^a\,$
with multiple Goldstone bosons ($\pi^a$) via 
the equivalence theorem\,\cite{ET}.
In the nonlinear realization, 
the bare mass-terms for quarks/leptons and Majorana neutrinos
result in contact interactions between these fermions and
Goldstones, of the type \,$f$-$\fbar$-$\pi^n$\, and
\,$\nL$-$\nL$-$\pi^n$\, ($n\geqq 1$) 
whose couplings are proportional 
to the corresponding fermion mass \,$m_{f(\nu)}^{~}$.\,
We classify the relevant contributions in Fig.\,1.
The contact diagram  Fig.\,1(a) gives a universal  
leading amplitude of \,${\mathcal O}(m_{f,\nu}^{~}E/v^n)$\, according to
power counting\,\cite{PC}, while the non-contact diagrams 
in Fig.\,1(b) are all subleading in energy.     
The contributions in Fig.\,1(c,d) involve the Goldstone
self-interactions and the new particle(s) generating the EWSB
(illustrated here by the SM Higgs boson $h^0$) which are 
model-dependent. A quantitative analysis of Fig.\,1(a,d)   
showed\,\cite{DH} that they may enhance the unitarity
bound by a factor of \,$[{\mathcal O}(2-3)]^{\f{1}{n-1}}$\, 
($n\geqq 3$) which is very close to one for large $n$\,.\, 
Thus it is justified to use the universal Fig.\,1(a) as a 
model-independent, conservative estimate for the unitarity bound.

Since power counting shows the leading amplitude of
$2\to n$ scattering to be
\,${\mathcal O}(1)(m_{f,\nu}^{~})^{2-\d}s^{\d/2}/v^n$\,
for $n\geqq 2$, 
we deduce the generical form of the squared amplitude,
\beq
\label{eq:T}
|\T|^2\,=\,
c_0^{~}(\theta_j)(2N_{c})^{2-\d}\,
m_{f,\nu}^{2(2-\d)}s^{\d}/v^{2n}\,,
\eeq
where \,$\d = 1\,(2)$\, for 
\,$f\fbar,\,\nL\nL,\,(V_L^{a_1}V_L^{a_2})$ $\to nV_L^a$\,
in the spin/color-singlet channel, and
$c_{0}^{~}(\theta_j)$ is a dimensionless coefficient with possible 
angular dependence. 
Using (\ref{eq:T}) we will compute the corresponding total cross
section $\sigma_{\rm inel}[2\to n]$ 
and determine its unitarity violation
scale from the condition (\ref{eq:UC2n}).

We observe that the key for resolving the puzzle posed in 
Eqs.\,(\ref{eq:UBfscot}) and (\ref{eq:KC}) is to analyze the  
additional $n$-dependent {\it dimensionless} factors in the 
exact n-body phase space integration; 
these factors will sufficiently
suppress the $E$-power enhancement $s^{n-2}$ in
Eq.\,(\ref{eq:UBfscot}).
Using (\ref{eq:T}), and
including the exact $n$-body phase space for the cross section
computation, we can formally derive the $2\to n$ unitarity limit, 
\beqa
\label{eq:UBus}
\dis
E^\star &=& v\[C_0\, 2^{4n-2}\pi^{2(n-1)}
            \varrho\, (n-1)!\,(n-2)!
\(2N_c\)^{\d-2}  
\right. \nonumber
\\
&& \hspace*{7mm} 
\left. \dis \times
\({v}/{\,m_{f,\nu}^{~}\,}\)^{2(2-\d)}\]^{\f{1}{2(n-2+\d)}}
%
\eeqa
where $\varrho$ is a symmetry factor for 
identical particles in the final state.
The factor $\,C_0$\, originates from the 
coefficient $c_0^{~}(\theta_j)$ whose possible angular
dependence is included in the exact $n$-body phase 
space integration. 
For the leading contact 
contributions in Fig.\,1(a) no such angular dependence exists. 
Since all large $n$ factors are explicitly
counted in (\ref{eq:UBus}),
we expect that, independent of detail, 
$~C_0^{\f{1}{\,2(n-2+\d)\,}} ~\sim~ 1~$ 
becomes increasingly more accurate as $n$ gets larger.
Thus, setting 
$\,\( C_0\,\varrho\)^{1/2(n-2+\d )}\approx 1$\, in (\ref{eq:UBus})
for simplicity, we derive an estimated unitarity limit \,$\Eu$\,
which is shown in Fig.\,2 (where $\,m_{\nu}^{~}=0.05$\,eV\, is chosen).   
Using the Stirling formula 
$\,n!\simeq n^ne^{-n}\sqrt{2\pi n}\,$, we deduce the
correct asymptotic behavior from (\ref{eq:UBus}),
\beq
\label{eq:asymp}
\dis
\Eu \,\to\, v\f{4\pi n}{e} ~~>~~ v\f{n}{3} \,,
~~~~~~~~({\rm for}~n\gg 1)\,, 
\eeq
which {\it agrees} with our kinematic condition (\ref{eq:KC}).

\begin{figure}[H]
\label{fig:Fig2}
\vspace*{-7mm}
\begin{center}
\includegraphics[width=8.5cm,height=6.8cm]{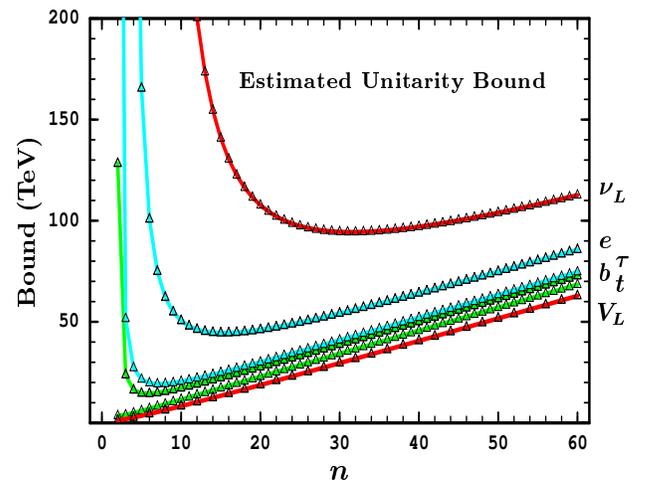}    
\vspace*{-5mm}
\caption{Realistic estimates of the unitarity limit $E^\star$,
from the scatterings 
\,$V_L^{a_1}V_L^{a_2},\,t\bar{t},\,b\bar{b},\,
   \tau^-\tau^+,\,e^-e^+,\,\nL\nL
   \to nV_L^a$\, (curves from bottom to top),
as a function of integer \,$n\,(\geqq 2)$\,.\,
}
\vspace*{-6mm}
\end{center}
\end{figure}

Fig.\,2 shows that for \,$V_L^{a_1}V_L^{a_2}$\, 
and \,$t\bar{t}$\, initial states
the strongest bound (minimum of the curve) still occurs at $n= n_s=2$;
while for {\it all light fermions including Majorana neutrinos}
the best limit for the scale of mass generation lies at a {\it new minimum}
with  $\,n= n_s > 2 \,$ and $\,\Eu\,$ no higher than about $100\,$TeV.
These limits are substantially tighter than 
the corresponding classic $2\to 2$ bounds.
The more precise calculations below 
will show these estimates are accurate to within a factor of two.
We stress that it is the {\it competition} between the
large asymptotic linear growth (\ref{eq:asymp}) and the strong
power suppression (\ref{eq:UBfscot}) that generates
a genuine {\it new minimum scale $E^\star_{\min}$} 
for all light fermions
at $\,n= n_s>2\,$, {\it independent of the EWSB scale.}

\vspace*{5mm}
\noindent  
\under{\it Scales of Mass Generation for Quarks and Leptons~}
\vspace*{3mm}

To improve the estimates in Fig.\,2, we will 
precisely compute the dimensionless coefficient $c_0^{~}$
in the leading squared-amplitude (\ref{eq:T}), so we can
determine the constant $C_0$ in Eq.\,(\ref{eq:UBus}) 
for each given \,$f\fbar ,\nL\nL\to nV_L^a$\, process 
(where the possible identical 
particle factor $\varrho$ will be explicitly included).
We first consider a pair of SM Dirac fermions $(f,f')$ which form a 
left-handed $SU(2)_L$ doublet $F_L=(f_L,f'_L)^T$ and two right-handed 
weak singlets $f_R$ and $f_R'$\,.\, We can formulate their bare Dirac 
mass-terms  $\,-m_f^{~}\ov{f}f - m_{f'}^{~}\ov{f'}f'\,$ 
into the gauge-invariant nonlinear form,
\beq
\label{eq:f-mass}
{\cal L}_f = \dis
-m_f\ov{F_L}U\(\ba{c} 1 \\ 0 \ea\)f_R
-m_{f'}\ov{F_L}U\(\ba{c} 0 \\ 1 \ea\)f'_R + {\rm H.c.}
\eeq 
where \,$U\,=\,\exp[i\pi^{a}\tau^{a}/v]$\,.\,
This gives $\,f$-$\fbar$-$\pi^n$ ($n\geqq 1$) 
contact interactions between fermions and Goldstone bosons.
We compute the leading scattering amplitude for 
$\,|{\rm in}\rangle \to (\pi^+)^k(\pi^-)^{\ell}(\pi^0)^{n-k-\ell}\,$,\,
where the initial state $|{\rm in}\rangle$ consists of two fermions
in the color-singlet channel\,\cite{DH}.\,

\begin{figure}[H]
\label{fig:Fig3}
\begin{center}
\vspace*{-6mm}
\hspace*{-1.9mm}
\includegraphics[width=8.9cm,height=7.7cm]{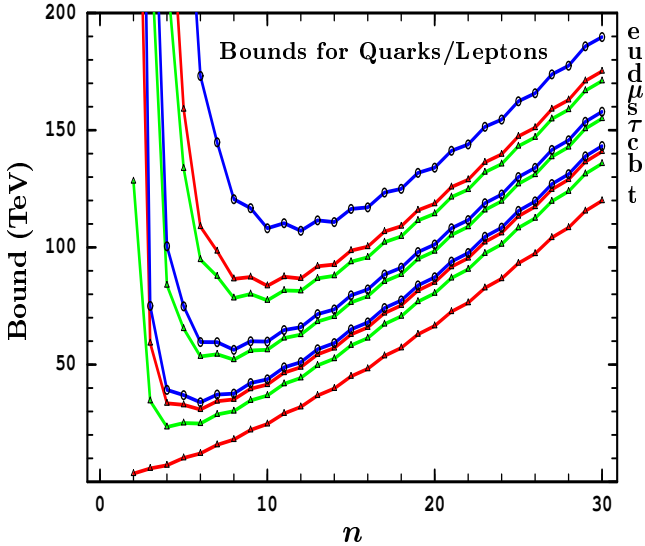} 
\hspace*{5mm}
\vspace*{-9mm}
\caption{Precise unitarity bounds $E^\star_f$ 
from scatterings 
$\,\xi_1^{~}\xi_2^{~}\to n\pi^a~(nV_L^a)$\,
as a function of \,$n$\,,\, where
$\,\xi_1^{~}\xi_2^{~}=t\bar{t},\,b\bar{b},\,c\bar{c},$ 
$ \,\tau^-\tau^+,\,
   s\bar{s},\,\mu^-\mu^+,\,d\bar{d},\,u\bar{u},\,e^-e^+
$\, 
for $n=$~even and
$\,\xi_1^{~}\xi_2^{~}=t_+\bar{b}_+,\,t_-\bar{b}_-,\,c_+\bar{s}_+,\,
   \tau_-^+\nu_-,\, c_-\bar{s}_-,\,\mu_-^+\nu_-,\,u_-\bar{d}_-,\,
    u_+\bar{d}_+,\,e^+_-\nu_- 
$\, for $n=$~odd, (curves from bottom to top).
The subscripts $\pm$ refer to helicity.
}
\end{center}
\vspace*{-3mm}
\end{figure}

Imposing unitarity on the $s$-partial wave, we derive 
corresponding unitarity bound for quarks and leptons,
\beqa
\label{eq:UBf-final}
E^\star_f ~~&=&~~ v\[\(\f{v}{m_{\widehat{f}}^{~}}\)^2
               \f{4\pi}{\,N_c\,\widehat{\R}_k^{\max}\,}
               \]^{\f{1}{2(n-1)}}  ,
\eeqa
where \,$\widehat{f}\in (f,\,f')$,\, and
$\,\widehat{\R}^{\max}_k \in 
         \(2\R_1^{\max},\,\R_2^{\max}\)\,$ 
with
\beq
\ba{lcl}
\R_1^{\max} &=&
\dis\f{\(\f{n}{2}!\)^2
}{~2^{3n-4}\pi^{2n-3}\,(n!)^2\,(n-1)!\,(n-2)!~} \,,
\\[5mm]
\R_2^{\max} &=&
\dis\f{\(\f{n-1}{2}\)!\(\f{n+1}{2}\)!}
{~2^{3n-4}\pi^{2n-3}\,(n!)^2(n-1)!\,(n-2)!~} \,,
\ea
\eeq
for $\,n=\,$(even,\,odd), respectively. 
We depict the bound $\,E_f^{\star}\,$ in Fig.\,3,
which shows that the best limits for all light fermions
occur at a {\it new minimum} $\,n= n_s>2\,$
with $\,E^\star_b=24.5\,$TeV for $b$ quark and 
     $\,E^\star_e=107\,$TeV for electron. 
These two limits only differ by a factor of \,$\sim\! 4$\,.

\begin{figure}[H]
\label{fig:Fig4}
\vspace*{-5mm}
\begin{center}
\includegraphics[width=8.5cm,height=7cm]{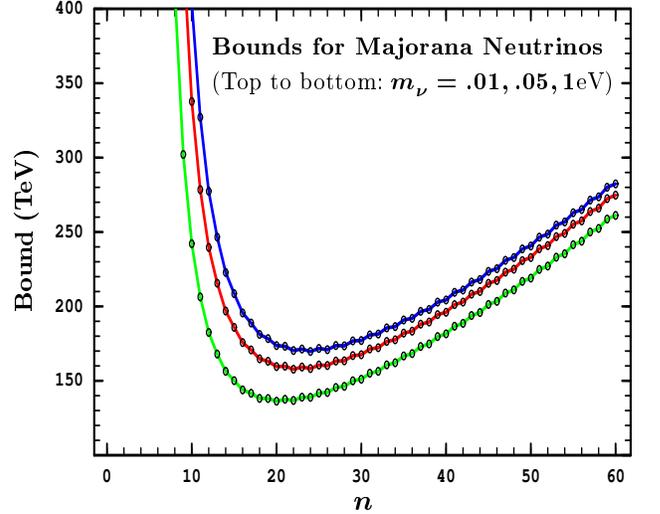}
\vspace*{-4.5mm}
\caption{Scales of mass generation for Majorana neutrinos:
the precise unitarity bounds \,$E^\star_\nu$\,
are plotted as a function of \,$n\,(\geqq 2)$\,.\,  
Only integer values of $n$ have a physical meaning.
}
\end{center}
\vspace*{-6mm}
\end{figure}

\begin{table*}
\vspace*{-4mm}
\label{Tab:Tab1}
\caption{  
Summary of the {\it strongest} unitarity limit 
\,$E^{\star\min}_{2\to n}$\, 
for each scattering $\,\xi^{~}_1\xi^{~}_2\to nV_L^a$\, 
(\,$\xi^{~}_{1,2} = V_L,\,f,\,\fbar,\,\nL$)
and the corresponding number of final state particles 
\,$n=n^{~}_s$\,,\, 
in comparison to the classic $2\to 2$ limit 
\,$E^{\star}_{2\to 2}$\,.
}
\vspace*{-1mm}
\begin{center}
\begin{tabular}{c||c|cccccc|ccc|c}
\hline\hline
&&&&&&&&&&&\\[-2.5mm]
~$\xi_1^{~}\xi_2^{~}$~      
& ~$V_L^{a_1}V_L^{a_2}$~     & ~$t\ov{t}$  &   
$b\ov{b}$       & $c\ov{c}$     &  $s\ov{s}$ &
$d\bar{d}$      & $u\ov{u}$    ~&~ $\tau^-\tau^+$~  & 
~$\mu^-\mu^+$~  & $e^-e^+$     & ~$\nL\nL$~~        
\\ [1.5mm] 
\hline\hline
&&&&&&&&&&&\\[-2.5mm]
~$M_W,m_{f,\nu}^{~}$\,(GeV)~
& 80.4 & 178  & 4.85 & 1.65 & 0.105 & 0.006 & 0.003 
& 1.777 & 0.106 
& $5.11\!\times\! 10^{-4}$ & $5\!\times\! 10^{-11}$
\\[1.5mm]
\hline
&&&&&&&&&&&\\[-2.5mm]
$n_s^{~}$ 
& $2$  & ~$2$ & $4$ & $6$ & $8$ & $10$ & $10$ ~& $6$ & $8$ 
& $12$ & ~$22$~~   
\\[1.2mm]
\hline
&&&&&&&&&&&\\[-2.5mm]
~$E^{\star(\min)}_{2\to n}$\,(TeV)~ 
& $1.2$ & ~\,$3.49$~ & ~$23.4$~ & \,$30.8$\, 
& \,$52.1$\, & \,$77.4$\, & \,$83.6$\,~
& $33.9$ & $56.3$ & $107$ & ~$158$~~   
\\[1.5mm]
\hline\hline
&&&&&&&&&&&\\[-2.5mm]
~$E^{\star}_{2\to 2}$\,(TeV)~ 
& $1.2$ & ~$3.49$ & $128$ & $377$ & $6\!\times\! 10^3$  & $10^5$ 
& $2\!\times\! 10^5$~ & $606$ & $10^4$ & ~$2\!\times\! 10^6$~ 
& ~~$1.1\!\times\! 10^{13}$~~   
\\[1.5mm]
\hline\hline
\end{tabular}
\end{center}
\vspace*{-3mm}
\end{table*}

\vspace*{3mm}
\noindent  
\under{\it Scales of Mass Generation for Majorana Neutrinos~}
\vspace*{3mm}

For Majorana neutrinos, we write their
dimension-3 bare mass term as
$\,-\f{1}{2}m_{\nu}^{ij}\nu_{Li}^T\widehat{C}
    \nu_{Lj}^{~}+{\rm H.c.},\,$
which takes the gauge-invariant nonlinear form 
\beq
\label{eq:nu-mass-pi}
{\cal L}_\nu ~=~ \dis
-\f{1}{2}m_\nu^{ij}{L^{\alpha}_i}^T \widehat{C}L^{\beta}_j
 \ov{\Phi}^{\alpha'}\ov{\Phi}^{\beta'}
\epsilon^{\alpha\alpha'} \epsilon^{\beta\beta'} + {\rm H.c.}
\,,
\eeq
where $\,\widehat{C}=i\gamma^2\gamma^0\,$,\,
$\,\ov{\Phi}=U(0,1)^T$\,,\, 
and $\,L_j=(\nu_{Lj}^{~}/\sqrt{2},\,\ell_{Lj})^T\,$.\,
We then quantitatively compute the high energy scattering
\,$\nL\nL \to   
   \(\pi^+\)^{\ell}\(\pi^-\)^{\ell}\(\pi^0\)^{n-2\ell}$\,.\,
In the actual calculation we use the real Majorana 
field $\,\chi =(\nL +\nu_L^c)/\sqrt{2}\,$ 
and compute the amplitude 
for three cases: 
(a) $n({\rm even})=2\ell$\,;\,   
(b) $n({\rm even}) > 2\ell$\,;\,
(c) $n({\rm odd})\geqq 3$\,.\,
We derive the unitarity bound,
\beqa
\label{eq:UBnu-final}
E^\star_\nu ~~&=&~~ v\[\(\f{v}{m_{\nu j}^{~}}\)^2
                       \f{4\pi}{\,\R_{\nu k}^{\max}\,}
             \,\]^{\f{1}{2(n-1)}}  ,
\eeqa
where $\,\R_{\nu k}^{\max}$\, ($k=1,2,3$) is 
a function of $n$ for three cases above\,\cite{DH}.
From Fig.\,4, we find the optimal limits, 
\beq
\label{eq:nuB-min}
\hspace*{-3mm}
E^\star_{\min} = 136,158,170\,{\rm TeV}, 
~~\,{\rm at}~\, n=n_s = 20,22,24,
\eeq
for typical inputs \,$m_{\nu j}^{~} = 1.0,\, 0.05,\, 0.01$\,eV.\,
This agrees with our estimate in Fig.\,2 to within
a factor of \,$\sim\! 1.7$.

\vspace*{4mm}
\noindent  
\under{\it Conclusions~}
\vspace*{1mm}

In this work, we have systematically 
analyzed the unitarity limits
from $2\to n$ inelastic scattering; 
this provides a universal upper bound 
on the scales of mass generation for
quarks, leptons and Majorana neutrinos. 
Table\,1 summarizes our results in 
comparison with classic $2\to 2$ limits\,\cite{AC,scott}.
It shows that the scattering
$\,f\bar{f}\to nV_L^a\,$ ($n\geqq 2$) does reveal
a {\it separate scale} for fermion mass generation.
Our new limits from  $2\to n$ scattering with
$\,n>2\,$ establish {\it new scales of mass generation} for all
light fermions including Majorana neutrinos, and 
are substantially stronger than the classic $2\to 2$ limits.
These new $2\to n$ bounds vary within the range of \,$3-160$\,TeV 
when the input fermion masses change from 
$m_t^{~}$ to $m_{\nu}^{~}$, 
as listed in Table\,1 where the masses are pole-masses except
that for $m_{u,d,s}^{~}$ the more precise $\ov{\rm MS}$ values are 
used\,\cite{PDG}.

In particular, for Majorana neutrinos with typical
masses $\,m_{\nu}^{~}=1.0-0.01\,$eV, 
the best upper limits on the
scale of mass generation fall in the range $136-170$\,TeV
(with $n= 20-24$). This is only a factor \,$\lesssim 7$\,
weaker than the lowest bound for light Dirac fermions 
despite their huge mass hierarchy
$\,m_{\nu}^{~}/m_{f\neq t}^{~} \leqq
m_{\nu}^{~}/m_b^{~}\approx 2\times(10^{-10}-10^{-12})\,$.
Hence, these limits are insensitive to the variation of 
fermion masses. 
Such a strong non-decoupling feature for the scale of new physics 
associated with light fermion mass generation is essentially due to
the {\it chiral structure} of fermion bare mass-terms, i.e., the 
fact that all left-handed SM fermions are weak-doublets but their
right-handed chiral partners are weak singlets (or possibly absent 
for a radiatively generated Majorana neutrino mass).
It is this feature that
makes the coupling strength of fermions to multiple Goldstone 
bosons (or effectively, multiple longitudinal gauge bosons)
{\it proportional to the fermion mass}, 
so the decoupling theorem\,\cite{DT} does not apply. 
For Majorana neutrino masses generated via the usual 
seesaw\,\cite{nu-seesaw} or radiative\,\cite{nu-rad} mechanism, 
this means that our new limits constrain the scale
of the leptonic Higgs Yukawa interaction, which must invoke
extra new fields (such as right-handed neutrinos or Zee-scalars
or triplet Higgs) as needed for generating lepton number violation 
and ensuring renormalizability 
(although our new bounds do not directly constrain
the masses of these new fields themselves 
which obey the decoupling theorem).

Finally, we have further estimated 
the \,$2\to n$\, unitarity limit on 
the EWSB scale via 
$\,V_L^{a_1} V_L^{a_2} \to n V_L^a$ 
($\pi^{a_1} \pi^{a_2} \to n \pi^a$)
with \,$n\geqq 2$,\, and find
that the best limit   
remains  \,$E^{\star}_W \simeq 1.2$\,TeV 
with $\,n=n_s=2\,$.
Also, the current study of \,$f\fbar \to nV_L^a\,(n\pi^a)$\,
scattering assumes $V_L^a$ ($\pi^a$) 
to be local field or remain local up to the limit 
$E^{\star}_{f}$.
If $V_L^a$ ($\pi^a$) becomes composite much below 
$E^{\star}_{f}$, the pure fermion process
$\,f\fbar \to (f\fbar)^n\,$
may be useful for a model-independent analysis.
A systematic expansion of this Letter 
is given in \cite{DH}.

\vspace*{1mm}
\noindent
{\bf Acknowledgments}~~
We thank S. Willenbrock for valuable discussions.
This work was supported by U.S. DOE 
under grant DE-FG03-93ER40757.

%


\begin{thebibliography}{0}

\bibitem{superK}
Y. Ashie, {\it et al.}, 
[SuperK-Kamiokande Collaboration],
Phys. Rev. Lett. {\bf 93}, 101801 (2004) 
[{\tt hep-ex/0404034}];
O. Elgaroy and O. Lahav, 
New J.\ Phys.\ {\bf 7}, 61 (2005).


\bibitem{wein5}
S. Weinberg, Phys. Rev. Lett. {\bf 43}, 1566 (1979).  


\bibitem{nu-seesaw}
P. Minkowski, Phys. Lett. B\,{\bf 67}, 421 (1977); 
M. Gell-Mann, P. Ramond, R. Slansky,
in {\it Proceedings of Workshop on Supergravity}, 
eds. F. van Nieuwenhuizen and D. Freedman (Amsterdam, 1979), 
p.315;
T. Yanagida, in {\it Proceedings of Workshop on Unified Theories
and Baryon Number in Universe}, 
eds. O. Sawada and A. Sugamoto (KEK, 1979), 
p.95;
S. L. Glashow, in {\it Quark and Leptons,} 
eds. M. Levy et al.\
(Plenum, NY, 1980);
R. N. Mohapatra and G. Senjanovic, 
Phys. Rev. Lett. {\bf 44}, 912 (1980).


\bibitem{nu-rad}
A. Zee, Phys. Lett. B\,{\bf 93}, 389 (1980);
L. Wolfenstein, Nucl. Phys. B\,{\bf 175}, 92 (1980);
S. T. Petcov, Phys. Lett.   B\,{\bf 115}, 401 (1982).



\bibitem{CCWZ}
C. G. Callen, S. Coleman, J. Wess, and B. Zumino,
Phys. Rev. {\bf 177}, 2247 (1969).


\bibitem{DM}
D. A. Dicus and V. S. Mathur, Phys. Rev. D\,{\bf 7}, 3111 (1973);
B. W. Lee, C. Quigg and H. B. Thacker, 
Phys. Rev. D\,{\bf 16}, 1519 (1977);
M. Veltman,  Acta. Phys. Polon. B\,{\bf 8}, 475 (1977).


\bibitem{AC}
T. Appelquist and M. S. Chanowitz, 
Phys. Rev. Lett. {\bf 59}, 2405 (1987).


\bibitem{MVW}
W. Marciano, G. Valencia, and S. Willenbrock,
Phys. Rev. D\,{\bf 40}, 1725 (1989).


\bibitem{scott-PRL}
F. Maltoni, J. M. Niczyporuk, and S. Willenbrock,
Phys. Rev. Lett. {\bf 86}, 212 (2001)
[{\tt hep-ph/0006358}].


\bibitem{scott}
F. Maltoni, J. M. Niczyporuk, and S. Willenbrock,
Phys. Rev. D\,{\bf 65}, 033004 (2002)
[{\tt hep-ph/0106281}].


\bibitem{ET}
For a comprehensive review,
H.-J. He, Y.-P. Kuang and C.-P. Yuan,
DESY-97-056 [{\tt hep-ph/9704276}].



\bibitem{PC}
H.-J. He, Y.-P. Kuang and C.-P. Yuan,
Phys. Rev.  D\,{\bf 55}, 3038 (1997); 
Phys. Lett. B\,{\bf 382}, 149 (1996). 


\bibitem{DH} 
D.\,A. Dicus and H.-J. He, Phys. Rev. D (2005), in Press,
[{\tt hep-ph/0409131}];
and also, H.-J. He and D. A. Dicus,
presentation at DPF-2004, {\tt hep-ph/0411024}. 

\bibitem{PDG}
Particle Data Group, Phys. Lett. B\,{\bf 592}, 1 (2004)
and {http://pdg.lbl.gov}.


\bibitem{DT}
T. Appelquist and J. Carrazone, 
Phys. Rev. D\,{\bf 11}, 2856 (1975).    




\end{thebibliography}
\end{document}